\documentclass[twocolumn,letter]{aastex63}

\include{multirow}

\newcommand{\tess}{{\it TESS}}

\newcommand{\ngts}{{NGTS}}

\newcommand{\coralie}{{CORALIE}}
\newcommand{\harps}{{HARPS}}
\newcommand{\feros}{{FEROS}}

\newcommand{\kms}{km\,s$^{-1}$}
\newcommand{\ms}{m\,s$^{-1}$}
\newcommand{\mss}{\mbox{m\,s$^{-2}$}}
\newcommand{\masy}{mas\,yr$^{-1}$}
\newcommand{\mpl}{\mbox{$M_{p}$}}
\newcommand{\rpl}{\mbox{$R_{p}$}}
\newcommand{\mstar}{\mbox{$M_{\star}$}}
\newcommand{\rstar}{\mbox{$R_{\star}$}}
\newcommand{\mjup}{\mbox{$M_{\rm Jup}$}}
\newcommand{\rjup}{\mbox{$R_{\rm Jup}$}}
\newcommand{\msun}{\mbox{$M_{\odot}$}}
\newcommand{\rsun}{\mbox{$R_{\odot}$}}

\newcommand{\vsini}{$v\sin{i}$}
\newcommand{\teff}{$T_{\rm eff}$}
\newcommand{\feh}{\mbox{$\rm [Fe/H]$}}
\newcommand{\logg}{$\log g$}

\newcommand{\Tstar}{NGTS-11}
\newcommand{\Tplanet}{NGTS-11\,b}

\newcommand{\TTplanet}{TOI-1847\,b}

\newcommand{\TICstar}{TIC-54002556}

\newcommand{\Tstarra}{$01^{\rm h}34^{'}05.15^{"}$}
\newcommand{\Tstardec}{$-14^{\circ}25' 08.9^{"}$}

\newcommand{\TGAIAGmag}{$12.1895 \pm 0.0002$}
\newcommand{\TGAIABPmag}{$12.6725 \pm 0.0012$}
\newcommand{\TGAIARPmag}{$11.5707 \pm 0.0011$}
\newcommand{\TGAIAPMRA}{$11.168 \pm 0.099$}
\newcommand{\TGAIAPMDec}{$14.047 \pm 0.049$}
\newcommand{\TGAIAPMplx}{$5.223 \pm  0.048$}

\newcommand{\TESSTmag}{$11.624 \pm 0.006$}
\newcommand{\TESSTmagshort}{\mbox{$11.62$}}
\newcommand{\APASSBmag}{$13.308 \pm 0.035$}
\newcommand{\APASSVmag}{$12.456 \pm 0.080$}
\newcommand{\APASSgmag}{$12.852 \pm 0.038$}
\newcommand{\APASSrmag}{$12.127 \pm 0.012$}
\newcommand{\APASSimag}{$11.864 \pm 0.083$ }

\newcommand{\MASSJ}{$10.855 \pm 0.025 $ }
\newcommand{\MASSH}{$10.401 \pm 0.023$ }
\newcommand{\MASSK}{$10.315  \pm 0.025$ }

\newcommand{\Tteff}{\mbox{$5050 \pm 80$}}%teff
\newcommand{\Tmetal}{\mbox{$0.22 \pm 0.08$}} %HARPS
\newcommand{\Tlogg}{\mbox{$4.5 \pm 0.1$}} %HARPS 
\newcommand{\Trotation}{\mbox{$1.1 \pm 0.8$}}

\newcommand{\Tstarmass}{\mbox{$0.862 \pm 0.028$}} 
\newcommand{\Tstarradius}{\mbox{$0.832 \pm 0.013$}}
\newcommand{\Tstardensity}{\mbox{$1.496 \pm 0.085$}}
\newcommand{\Tstarage}{\mbox{$3.9 \pm 1.6$}}

\newcommand{\TfitTO}{\mbox{$ 2458390.7043 \pm 0.0016$}} 
\newcommand{\Tfitperiod}{\mbox{$35.45533 \pm 0.00019$}}
\newcommand{\Tperiodshort}{\mbox{$35.46$}} 
\newcommand{\TfitradiusI}{\mbox{$ 0.0183 \pm _{0.0025}^{0.0029}$}} 
\newcommand{\Tfitk}{\mbox{$0.1010 \pm _{0.0037}^{0.0030}$}} 
\newcommand{\Tfitb}{\mbox{$0.81 \pm _{0.10}^{0.13}$}} 
\newcommand{\TfithItess}{\mbox{$0.7406 \pm _{0.0030}^{0.0030}$}}
\newcommand{\TfithIItess}{\mbox{$ 0.427 \pm _{0.046}^{0.046}$}} 
\newcommand{\TfithIngts}{\mbox{$0.7144 \pm _{0.0030}^{0.0031}$}}
\newcommand{\TfithIIngts}{\mbox{$0.442 \pm _{0.046}^{0.043}$}}
\newcommand{\Tfitl}{\mbox{$0.051 \pm _{0.034}^{0.048}$}} 
\newcommand{\TfitK}{\mbox{$0.0212 \pm _{0.0041}^{0.0046}$}}
\newcommand{\Tfitfs}{\mbox{$0.13 \pm _{0.28}^{0.26}$}} 
\newcommand{\Tfitfc}{\mbox{$0.19 \pm _{0.22}^{0.16}$}} 
\newcommand{\TfitV}{\mbox{$21.4094 \pm _{0.0028}^{0.0029} $}} 
\newcommand{\TfitHF}{\mbox{$-0.042 \pm _{0.010}^{0.010}$}} 
\newcommand{\Tfite}{\mbox{$0.13 \pm_ {0.09}^{0.10}$}} 
\newcommand{\Tfitw}{\mbox{$32 \pm_ {76}^{53}$}} 
\newcommand{\Tfitdur}{\mbox{$3.59 \pm_ {0.37}^{0.18}$}} 

\newcommand{\TdelTshort}{\mbox{$390.0082$}}
\newcommand{\TdelT}{\mbox{$390.0082 \pm 0.0032$}}

\newcommand{\Tplanetmass}{\mbox{$0.344 \pm _{0.073}^{0.092}$}}
\newcommand{\Tplanetradius}{\mbox{$0.817 \pm _{0.032}^{0.028}$}} 
\newcommand{\Teq}{\mbox{$435 \pm _{32}^{34}$}}
\newcommand{\Tpgrav}{\mbox{$14.1 \pm _{3.4}^{5.3}$}}
\newcommand{\TpH}{\mbox{$141 \pm _{51}^{74}$}}
\newcommand{\Tpa}{\mbox{$0.2010 \pm _{0.0022}^{0.0021}$}}
\newcommand{\Tpincl}{\mbox{$89.16 \pm _{0.29}^{0.20}$}}
\newcommand{\Tpden}{\mbox{$0.78 \pm _{0.17}^{0.21} $}} % g/cm^3

\received{June 1, 2019}
\revised{January 10, 2019}
\accepted{\today}
\submitjournal{ApJL}

\shorttitle{The warm Saturn \Tstar\,$\rm b$}
\shortauthors{Gill et al.}

\graphicspath{{./}{figures/}}
\begin{document}

\title{\Tplanet\ (\TTplanet):  A transiting warm Saturn recovered from a \tess\ single-transit event}
%\title{A transiting warm Saturn recovered from a \tess\ single-transit event}

% Ordered authors 

\correspondingauthor{Samuel Gill} 
\email{samuel.gill@warwick.ac.uk} 
\author[0000-0002-4259-0155]{Samuel Gill}
\affil{Centre for Exoplanets and Habitability, University of Warwick, Gibbet Hill Road, Coventry CV4 7AL, UK}
\affiliation{Department of Physics, University of Warwick, Gibbet Hill Road, Coventry CV4 7AL, UK}

\author[0000-0003-1452-2240]{Peter J.\ Wheatley} 
\affil{Centre for Exoplanets and Habitability, University of Warwick, Gibbet Hill Road, Coventry CV4 7AL, UK}
\affiliation{Department of Physics, University of Warwick, Gibbet Hill Road, Coventry CV4 7AL, UK}

\author[0000-0002-8824-9956]{Benjamin F. Cooke} 
\affiliation{Department of Physics, University of Warwick, Gibbet Hill Road, Coventry CV4 7AL, UK}

\author[0000-0002-5389-3944]{Andr{\'e}s Jord{\'a}n} 
\affiliation{Facultad de Ingenier\'ia y Ciencias, Universidad Adolfo Ib\'a\~{n}ez, Av.Diagonal las Torres 2640, Pe\~{n}alol\'en, Santiago, Chile}
\affiliation{Millennium Institute for Astrophysics, Chile}

\author[0000-0002-5254-2499]{Louise~D.~Nielsen} 
%\affiliation{Observatoire de Gen{\`e}ve, Universit{\'e} de Gen{\`e}ve, 51 Ch. des Maillettes, 1290 Sauverny, Switzerland}
\affiliation{Observatoire astronomique de l'Universit\'e de Gen\`eve, Chemin des maillettes 51, 1290 Sauverny, Switzerland}

\author[0000-0001-6023-1335]{Daniel Bayliss}
\affiliation{Department of Physics, University of Warwick, Gibbet Hill Road, Coventry CV4 7AL, UK}

\author[0000-0001-7416-7522]{David~R.~Anderson}
\affiliation{Department of Physics, University of Warwick, Gibbet Hill Road, Coventry CV4 7AL, UK}

\author[0000-0002-1896-2377]{Jose I. Vines} 
\affiliation{Departamento de Astronom\'ia, Universidad de Chile, Camino El Observatorio 1515, Las Condes, Santiago, Chile}

\author[0000-0001-9699-1459]{Monika~Lendl}
\affiliation{Observatoire astronomique de l'Universit\'e de Gen\`eve, Chemin des maillettes 51, 1290 Sauverny, Switzerland}
\affiliation{Space Research Institute, Austrian Academy of Sciences, Schmiedlstr. 6, 8042 Graz, Austria}

% alphabetical Authors

\author{Jack S. Acton}
\affiliation{School of Physics and Astronomy, University of Leicester, University Road, Leicester LE1 7RH,  UK}

\author[0000-0002-5080-4117]{David J. Armstrong}
%\altaffiliation{STFC Ernest Rutherford Fellow}
\affil{Centre for Exoplanets and Habitability, University of Warwick, Gibbet Hill Road, Coventry CV4 7AL, UK}
\affil{Department of Physics, University of Warwick, Gibbet Hill Road, Coventry CV4 7AL, UK}

\author{Fran\c{c}ois Bouchy} 
\affiliation{Observatoire astronomique de l'Universit\'e de Gen\`eve, Chemin des maillettes 51, 1290 Sauverny, Switzerland}

\author{Rafael Brahm} %(rafael.brahm@uai.cl) 
\affiliation{Facultad de Ingenier\'ia y Ciencias, Universidad Adolfo Ib\'a\~{n}ez, Av.Diagonal las Torres 2640, Pe\~{n}alol\'en, Santiago, Chile}
\affiliation{Millennium Institute for Astrophysics, Chile}

\author{Edward M. Bryant}
\affil{Centre for Exoplanets and Habitability, University of Warwick, Gibbet Hill Road, Coventry CV4 7AL, UK}
\affiliation{Department of Physics, University of Warwick, Gibbet Hill Road, Coventry CV4 7AL, UK}

\author[0000-0003-0684-7803]{Matthew R. Burleigh}
\affiliation{School of Physics and Astronomy, University of Leicester, University Road, Leicester LE1 7RH,  UK}

\author[0000-0003-2478-0120]{Sarah L. Casewell}
\affiliation{School of Physics and Astronomy, University of Leicester, University Road, Leicester LE1 7RH,  UK}

\author{Philipp Eigm\"uller}
\affiliation{Institute of Planetary Research, German Aerospace Center, Rutherfordstrasse 2, 12489, Berlin, Germany}

\author[0000-0001-9513-1449]{N{\'e}stor Espinoza} %(nespinoza@stsci.edu) 
\affiliation{Space Telescope Science  Institute, 3700 San Martin Drive, Baltimore, MD 21218, USA}

\author[0000-0003-2851-3070]{Edward Gillen}
\affiliation{Astrophysics Group, Cavendish Laboratory, J.J. Thomson Avenue, Cambridge CB3 0HE, UK}
\affiliation{Winton Fellow}

\author{Michael R.Goad} %(mg159@le.ac.uk)
\affiliation{School of Physics and Astronomy, University of Leicester, University Road, Leicester LE1 7RH,  UK}

\author{Nolan Grieves} 
%\affiliation{Observatoire de Gen{\`e}ve, Universit{\'e} de Gen{\`e}ve, 51 Ch. des Maillettes, 1290 Sauverny, Switzerland}
\affiliation{Observatoire astronomique de l'Universit\'e de Gen\`eve, Chemin des maillettes 51, 1290 Sauverny, Switzerland}

\author[0000-0002-3164-9086]{Maximilian~N.~G{\"u}nther}
\affiliation{Department of Physics and Kavli Institute for Astrophysics and Space Research, Massachusetts Institute of Technology, 70 Vassar Street, Cambridge, MA 02139,
USA}
\affiliation{Juan Carlos Torres Fellow}

\author[0000-0002-1493-300X]{Thomas Henning} %(henning@mpia.de) 
\affiliation{Max-Planck-Institut f\"ur Astronomie, K\"onigstuhl 17, 69117 Heidelberg, Germany}

\author[0000-0002-5945-7975]{Melissa J. Hobson} %(melihobson@gmail.com) 
\affiliation{Instituto de Astrof\'isica, Facultad de F\'isica, Pontificia Universidad Cat\'olica de Chile}
\affiliation{Millennium Institute for Astrophysics, Chile}

\author{Aleisha Hogan}
\affiliation{School of Physics and Astronomy, University of Leicester, University Road, Leicester LE1 7RH,  UK}

\author{James S. Jenkins}
\affiliation{Departamento de Astronom\'ia, Universidad de Chile, Camino El Observatorio 1515, Las Condes, Santiago, Chile}
\affiliation{Centro de Astrof\'isica y Tecnolog\'ias Afines (CATA), Casilla 36-D, Santiago, Chile}

\author{James~McCormac}
\affiliation{Department of Physics, University of Warwick, Gibbet Hill Road, Coventry CV4 7AL, UK}

\author{Maximiliano Moyano}
\affiliation{Instituto de Astronom\'ia, Universidad Cat\'olica del Norte, Angamos 0610, 1270709 Antofagasta, Chile}

\author[0000-0002-4047-4724]{Hugh P. Osborn}
\affiliation{NCCR/PlanetS, Centre for Space \& Habitability, University of Bern, Bern, Switzerland}
\affiliation{Department of Physics and Kavli Institute for Astrophysics and Space Research, Massachusetts Institute of Technology, 70 Vassar Street, Cambridge, MA 02139,
USA}

\author[0000-0001-9850-9697]{Don Pollacco}
\affil{Centre for Exoplanets and Habitability, University of Warwick, Gibbet Hill Road, Coventry CV4 7AL, UK}
\affil{Department of Physics, University of Warwick, Gibbet Hill Road, Coventry CV4 7AL, UK}

\author[0000-0002-3012-0316]{Didier Queloz}
\affiliation{Astrophysics Group, Cavendish Laboratory, J.J. Thomson Avenue, Cambridge CB3 0HE, UK}

\author{Heike Rauer}
\affiliation{Institute of Planetary Research, German Aerospace Center, Rutherfordstrasse 2, 12489, Berlin, Germany}

\author{Liam Raynard}
\affiliation{School of Physics and Astronomy, University of Leicester, University Road, Leicester LE1 7RH,  UK}

\author{Felipe Rojas} % (firojas@uc.cl) 
\affiliation{Instituto de Astrof\'isica, Facultad de F\'isica, Pontificia Universidad Cat\'olica de Chile}
\affiliation{Millennium Institute for Astrophysics, Chile}

\author{Paula Sarkis} %(sarkis@mpia.de) 
\affiliation{Max-Planck-Institut f\"ur Astronomie, K\"onigstuhl 17, 69117 Heidelberg, Germany}

\author[0000-0002-2386-4341]{Alexis~M.~S.~Smith}
\affiliation{Institute of Planetary Research, German Aerospace Center, Rutherfordstrasse 2, 12489, Berlin, Germany}

\author{Marcelo Tala Pinto} %(mtala@lsw.uni-heidelberg.de) 
\affiliation{Landessternwarte, Zentrum f\"ur Astronomie der Universit\"at Heidelberg, K\"onigstuhl 12, 69117 Heidelberg, Germany}
\affiliation{Millennium Institute for Astrophysics, Chile}

\author{Rosanna H. Tilbrook}
\affiliation{School of Physics and Astronomy, University of Leicester, University Road, Leicester LE1 7RH,  UK}

\author[0000-0001-7576-6236]{St\'ephane Udry}
\affiliation{Observatoire astronomique de l'Universit\'e de Gen\`eve, Chemin des maillettes 51, 1290 Sauverny, Switzerland}

\author{Christopher A. Watson}
\affiliation{Astrophysics Research Centre, School of Mathematics and Physics, Queen's University Belfast, BT7 1NN, Belfast, UK}

\author[0000-0001-6604-5533]{Richard G. West}
\affiliation{Department of Physics, University of Warwick, Gibbet Hill Road, Coventry CV4 7AL, UK}

\nocollaboration{50}

\begin{abstract}

We report the discovery of \Tplanet\ (=\TTplanet), a transiting Saturn in a \Tperiodshort-day orbit around a mid K-type star (\teff=\Tteff\,K).  We initially identified the system from a single-transit event in a \tess\ full-frame image light-curve.  Following seventy-nine nights of photometric monitoring with an \ngts\ telescope, we observed a second full transit of \Tplanet\ approximately one year after the \tess\ single-transit event.  The \ngts\ transit confirmed the parameters of the transit signal and restricted the orbital period to a set of 13 discrete periods. We combined our transit detections with precise radial velocity measurements to determine the true orbital period and measure the mass of the planet.  We find \Tplanet\ has a radius of \Tplanetradius\,$\rjup$, a mass of \Tplanetmass\,$\mjup$, 
and an equilibrium temperature of just \Teq\,K, making it one of the coolest known transiting gas giants. \Tplanet\ is the first exoplanet to be discovered after being initially identified as a \tess\ single-transit event, and its discovery highlights the power of intense photometric monitoring in recovering longer-period transiting exoplanets from single-transit events.

\end{abstract}


\section{Introduction}\label{sec:intro}

\begin{figure}
    \centering
    \includegraphics[scale=0.5]{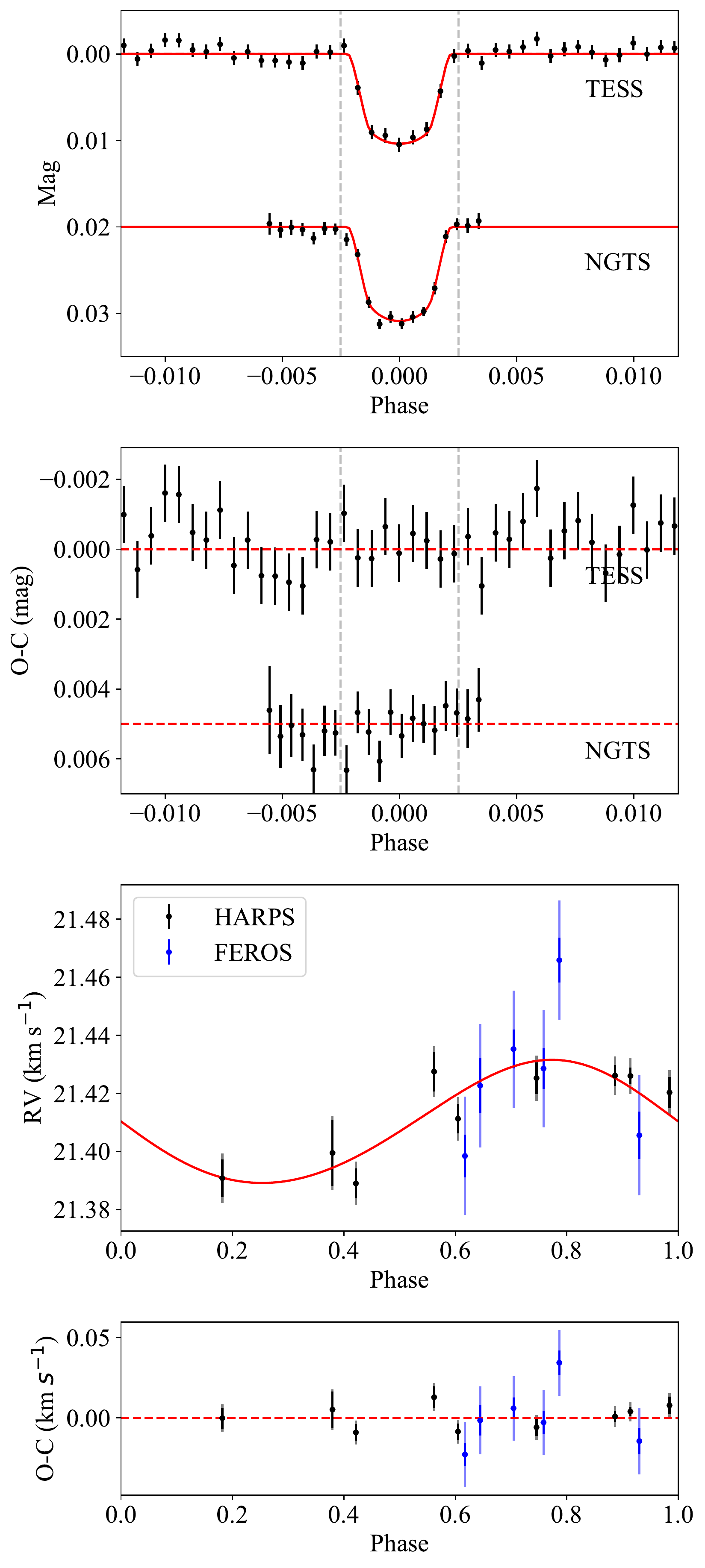}
    \caption{Our best-fitting model to the photometric and radial velocity observations of \Tstar. Upper panel shows the transit photometry of \tess\ and \ngts\ with the model in red.  Dashed vertical grey lines indicate start of ingress and end of egress. Photometric residuals are shown in the upper-middle panel.  \ngts\ photometry has been binned to 30\,min for plotting purposes, but was fitted at the full 13\,s time resolution. The lower panels show the radial velocity observations and residuals from \feros\ (blue) and \harps\ (black) along with our best-fitting model (red). Semi-transparent error bars represent uncertainties with jitter values added in quadrature.   }
    \label{fig:Figure_1}
\end{figure}

Wide-field photometric surveys have uncovered a large population of transiting exoplanets around bright stars \citep[e.g.][]{Bakos04,2006PASP..118.1407P,2010Sci...327..977B}, providing key opportunities to measure precise planetary radii and densities, constrain bulk composition, and to characterise planetary atmospheres \citep[e.g.][]{Seager03,Fortney07,Sing16}. However, the transit geometry imposes a strong selection bias for close-in orbits, and only a handful of well-characterised transiting exoplanets are known with orbital periods longer than about 30\,days. This limits our characterisation of the wider exoplanet population, as well as its evolution through planetary migration \citep[e.g.][]{2004ASPC..321..355N,2004ASPC..321..298J,2013A&A...558A.109A}. It also prevents us from using transits to characterise exoplanets in the habitable zone \citep{1993Icar..101..108K,2013ApJ...765..131K}, except around very late-type host stars \citep[e.g.][]{Gillon17}.

The extremely wide field of the \tess\ survey \citep{2015JATIS...1a4003R} is providing many new exoplanet candidates around bright stars. Longer-period planets can be confirmed at high ecliptic latitudes, where \tess\ monitors stars for up to one year \citep[e.g.][]{2020MNRAS.494..750E,2020AJ....159..241D}.  
%TOI 831b has 4 transits across multiple sectors and ecliptic latitude of -80 (only 1.5% of sky at higher lats)
%TOI 1456b has 2 transits from 2 sectors and ecliptic latitude of +50 (only 23% of sky at higher lats, or about 77% of sky at lower lats). 
However, with an observing baseline of only 27\,days across most of the sky, \tess\ typically detects only a single transit for longer-period planets \citep{2018A&A...619A.175C,2019AJ....157...84V}. 
Follow-up of these single transit events provide the opportunity to discover and characterise many more longer-period planets \citep{2008ApJ...688..616Y,Yao_2019}.
%Yee+Gaudi08 was about Kepler single-transit events, periods predicted to 20% (assuming circular), and RV follow up. 
%Yao+Pepper+19 was about precovery with KELT
However, a single-transit event provides only a weak constraint on the orbital period of the planet \citep{2016MNRAS.457.2273O,2019MNRAS.489.3149S}, and so follow-up observations designed to confirm the exoplanet are expensive: whether aiming to detect additional transits, or to detect the radial velocity variations of the host star.

The \ngts\ project \citep{2018MNRAS.475.4476W} has commenced a program to follow-up \tess\ single-transit event candidates, and has already discovered three longer-period, low-mass eclipsing binary systems: 
TIC-238855958 \citep{2020MNRAS.491.1548G},
TOI-222 \citep{2020MNRAS.492.1761L}, and
TIC-231005575 \citep{2020arXiv200209311G}.

In this letter we present the first discovery of an exoplanet based on a single-transit event from \tess. We show how an observing strategy focusing on detecting a second transit, in this case with NGTS, enables longer-period planets to be confirmed and characterised using a very modest number of radial velocity measurements.

%\begin{figure}
%    \centering
%    \includegraphics[scale=0.5]{Figure_1.pdf}
%    \caption{Our best-fitting model to the photometric and radial velocity observations of \TTstar/\Tstar. Upper panel shows the transit photometry of \tess\ and \ngts\ with the model in red.  Dashed vertical grey lines indicate start of ingress and end of egress. Photometric residuals are shown in the upper-middle panel.  \ngts\ photometry has been binned to 30\,min for plotting purposes, but was fitted at the full 13\,s time resolution. The lower panels show the radial velocity observations and residuals from \feros\ (blue) and \harps\ (black) along with our best-fitting model (red).}
%    \label{fig:Figure_1}
%\end{figure}

\section{Observations and analysis}
\subsection{\tess\ single transit detection}
\label{sec:tess}

Using light curves produced from the \tess\ calibrated full-frame images by \citet{2018AJ....156..132O}, 
we conducted a systematic search for single-transit events using the procedure described by \citet{2020MNRAS.491.1548G}.

The K-dwarf \TICstar\ (hereafter \Tstar, $T = $ \TESSTmagshort, properties in Table~\ref{tab:Table1}) was observed at 30-minute cadence with \tess\ Camera~1 during Sector~3 of the mission (2018 September 20 -- 2018 October 18). \Tstar\ was not selected for \tess\ 2-minute cadence observations, so does not have a light curve from the SPOC pipeline.

We identified a promising single-transit event from \Tstar\ approximately nine days into Sector~3, at JD\,2458390.7, with a transit width of 4\,h and depth of 1\,percent (shown in top panel of Fig.\,\ref{fig:Figure_1}). We inspected the individual \tess\ calibrated full-frame images for asteroids and any other anomalies. We also searched for blended objects in Gaia DR2 \citep{2018A&A...616A...1G} and checked for known nearby exoplanets or eclipsing binaries that might be the source of the transit event. We found no reason to believe this single-transit candidate was a false-positive.

Having identified \Tstar\ as hosting a viable exoplanet candidate, we created a higher quality \tess\ light curve from the \tess\ calibrated full-frame images by carefully selecting source and background pixel apertures in order to minimise blending with neighbouring objects. 
Background pixels were selected from a $15\times15$\,pixel median master frame using an iterative sigma-clipping process. Source pixels were selected from a central $7\times7$\,pixel region where median pixel counts exceeded the background level by 100 times the standard deviation in the background. To exclude source pixels contaminated by neighbouring objects, we also required a monotonic decrease in counts with distance from the target position. The final source aperture included 19 pixels.

These apertures better exclude fainter, nearby companions TIC-54002560 (48.51\arcsec; $T = 14.28$) and TIC-54002559 (56.41\arcsec; $T = 14.51$). Frames with a quality flag larger than 0 were rejected, and the light curve was detrended using a median filter of width 10.5\,h. The resulting \tess\ light curve around the time of the transit is shown in Fig.\,\ref{fig:Figure_1}, and this was used for the remainder of this work.

\subsection{A second transit with \ngts}\label{secondEpoch}

We used the \ngts\ facility \citep{2018MNRAS.475.4476W} located at the ESO Paranal Observatory in Chile to monitor \Tstar\ photometrically, searching for additional transits. \ngts\ was designed specifically for very high precision photometry of exoplanet transits, and it is thus well suited to photometric follow-up of \tess\ single-transit candidates \citep{2018MNRAS.475.4476W}. We started monitoring \Tstar\ with a single \ngts\ telescope on the night of 2019 August 11. We observed each subsequent night with 10\,s exposures whenever the target was above an elevation of 30\,deg and weather conditions allowed.  The images were reduced to light curves each day using the on-site \ngts\ real-time aperture photometry pipeline described by \citet{bryant:2020}.

We used the template matching algorithm described by \citet{2020MNRAS.491.1548G,2020arXiv200209311G} to 
search the \ngts\ light curve for additional transit events. This involved using a model fit to the \tess\ transit as a template, which was placed at each possible transit epoch within the \ngts\ light curve and the improvement in log-likelihood calculated ($\Delta \log \mathcal{L}= -\Delta \chi^2/2$). From empirical injection tests, we conservatively estimated that a significant template match would be consistent with $\Delta \log \mathcal{L} > 50$. We observed \Tstar\ for 79 nights (105\,642 exposures) before a second transit event was detected ($\Delta\log\mathcal{L} = 364$) on the UT night of 24 October 2019.  The transit event was centred at JD\,2458780.712, which is 390\,days after the initial \tess\ single-transit. The \ngts\ transit detection %of \TTplanet\ (=\Tplanet) 
is plotted in the top panel of Fig.\,\ref{fig:Figure_1}.

\subsection{Photometric constraints on orbital period}
\label{sec:aliases}

We measured the precise separation of the \tess\ and \ngts\ transit detections using a joint transit fit to both light curves, finding $\Delta T = \TdelT$\,days. The orbital period of the companion to \Tstar\ must be an integer fraction of this duration. We label these candidate periods as $P_{n} =(\TdelTshort/n)$ where $n$ is an integer. These candidate periods are indicated with vertical lines in Figure\,\ref{fig:Figure_2}.

\begin{table*}[t]
    \centering
    \caption{Catalogued, measured and derived properties of \Tstar\ and \Tplanet. 
    % Pete - I commented out the following cos its not needed I think
    %Asymmetric errors are reported in brackets and correspond to the difference between the median and the 16$^{th}$ (lower value) and 84$^{th}$ (upper value) percentile. Asymmetric errors which differed by a factor of 2 are reported in quadrature as symmetric uncertainties with $\pm$.
    }
    \begin{tabular}{cc|cc}
    \hline
    \hline
    \multicolumn{2}{l|}{Catalogue data} & \multicolumn{2}{l}{Model parameters}     \\
    \hline
%Gaia Source ID & \TGAIAid        & $T_0$ (BJD-TDB) & \TfitTO  \\
TIC ID & \TICstar        & $T_0$ (BJD-TDB) & \TfitTO  \\
R.A.  &  \Tstarra      & $P$ (d) &  \Tfitperiod \\
decl.  & \Tstardec     & \rstar / $a$ & \TfitradiusI\\
G & \TGAIAGmag                   & \rpl / \rstar & \Tfitk \\
BP & \TGAIABPmag                 & $b$          & \Tfitb \\
RP & \TGAIARPmag                &  $h_{1,\rm \tess}$  & \TfithItess  \\
pmRA (\masy)   & \TGAIAPMRA  & $h_{2,\rm \tess}$ & \TfithIItess \\
pmDec (\masy)  & \TGAIAPMDec  & $h_{1,\rm \ngts}$  & \TfithIngts   \\
Parallax ($\rm mas$) & \TGAIAPMplx  & $h_{2,\rm \ngts}$  & \TfithIIngts \\
\tess\ (T)   & \TESSTmag &  $l_{3,\tess}$ & \Tfitl\\
APASS9 (B) & \APASSBmag &  $K_\star$ (\kms) & \TfitK \\
APASS9 (V) & \APASSVmag & $\sqrt{e}\sin \omega$ & \Tfitfs  \\
APASS9 (g') & \APASSgmag & $\sqrt{e}\cos \omega$& \Tfitfc  \\
APASS9 (r') & \APASSrmag &  $\gamma$ (\kms)& \TfitV \\ 
APASS9 (i') &\APASSimag & $\Delta \gamma_{\rm FEROS}$ (\kms)& \TfitHF\\
\cline{3-4}
2MASS (J)  & \MASSJ  &  \multicolumn{2}{l}{Derived planet properties} \\
\cline{3-4}
2MASS (H)  & \MASSH  & \mpl\ (\mjup) & \Tplanetmass \\
2MASS (K$_{\rm s}$)& \MASSK &  \rpl\ (\rjup) &  \Tplanetradius  \\
\cline{1-2}
\multicolumn{2}{l|}{Stellar parameters} &  $a\,(\rm au)$ & \Tpa \\  
\cline{1-2}
\teff\ (K) &  \Tteff & e & \Tfite   \\
\feh\ ($\rm dex$) &  \Tmetal &  $\omega$ (deg) & \Tfitw \\  
$\log g_\star$ ($\rm dex$) &  \Tlogg &  $i$ (deg) & \Tpincl  \\
\vsini\ (\kms) & \Trotation  &  $T_{\rm dur}$ (hr) & \Tfitdur \\
\mstar\ (\msun) &  \Tstarmass  & $g_p$ (\mss)& \Tpgrav\\
\rstar\ (\rsun) & \Tstarradius  & $T_{\rm eq}$ $(\rm K)$ & \Teq  \\
$\rho_\star$\ ($\rho_{\sun}$) & \Tstardensity & $\rho_{p}$\ (g\,cm$^{-3}$) & \Tpden  \\
Age $(\rm Gyr)$ & \Tstarage & $H_p$ (km)& \TpH \\
\hline
    \end{tabular}
    \label{tab:Table1}
\end{table*}

All candidate periods beyond $P_{18}$ were ruled out by the \tess\ data, since additional transits would have been detected. Similarly, five candidate periods ($P_{10}$, $P_{13}$, $P_{14}$, $P_{15}$ and $P_{17}$) were ruled out by the \ngts\ photometric monitoring as well as additional observations on 2019 November 22 and 2019 December 1 (red dashed vertical lines in Figure\,\ref{fig:Figure_2}). This left just 13 candidate orbital periods (green dashed vertical lines in Figure\,\ref{fig:Figure_2}).

\subsection{Radial velocity measurements}
Following the NGTS transit detection, we immediately began radial-velocity follow-up using the \coralie\ fiber-fed \'{E}chelle spectrograph installed on the 1.2-m Leonard Euler telescope at the ESO La Silla Observatory \citep{2001A&A...379..279Q}. We made three $600$\,s observations of \Tstar\ over 54\,days (2019 October 29, November 28, and December 25), the first just 5\,days after the NGTS transit detection. The spectra were reduced using the \coralie\ standard reduction pipeline, with radial velocity measurements derived using the cross-correlation technique and a numerical G2 mask. The radial velocities had no variation above 80\,\ms\ showing that the transiting companion is unlikely to be a low-mass star.

This motivated an additional nine 1800\,s radial velocity measurements spanning 63\,days with the \harps\ spectrograph ($R=115000$) on the 3.6-m ESO telescope \citep{2002Msngr.110....9P} and six measurements spanning 11\,days with the \feros\ spectrograph ($R=48000$) on the MPG/ESO 2.2-m Telescope  \citep{kaufer:99}. The FEROS data were reduced with the CERES pipeline \citep{2017PASP..129c4002B}. These data are plotted in the lower panels of Figure~\ref{fig:Figure_1}.

\subsection{Stellar properties}
\label{sec:specpars}

We used our \harps\ spectra to determine the properties of the host star \Tstar. The individual spectra were shifted to the laboratory frame of reference and co-added to produce a combined spectrum with a signal-to-noise ratio of 44. We used \textsc{ispec} \citep{2014ASInC..11...85B} to synthesise models using the the radiative transfer code \textsc{spectrum} \citep{1999ascl.soft10002G}, MARCS model atmospheres \citep{2008A&A...486..951G}, and version 5 of the GAIA ESO survey (GES) atomic line list within \textsc{ispec} with solar abundances from \citet{2009ARA&A..47..481A}. 
%We synthesised models that best matched the observed spectrum, and estimated uncertainties by adjusting parameters until fits were unacceptable. 
We used the H$\alpha$, NaI\,D, and MgI\,b lines to determine the stellar effective temperature, \teff, and surface gravity, \logg.   We used individual FeI and FeII lines to determine metallicity, \feh, and the rotational broadening projected into the line of sight, \vsini. Values of macroturbulence and microturbulence were calculated using equations~5.10 and 3.1 respectively from \citet{2015PhDT........16D}.

We used the method described in \citet{2020MNRAS.491.1548G} to determine the mass, radius, and age of \Tstar.  This method uses Gaia magnitudes and parallax \citep{2018A&A...616A...1G} along with \teff\ and \feh\ from our spectroscopy to determine the best-fitting stellar parameters with respect to MESA models \citep{2016ApJS..222....8D,2016ApJ...823..102C}. 

The results of our analysis are presented in Table~\ref{tab:Table1}. We find that \Tstar\ is a mid K-dwarf star (\teff=\Tteff\,K) with super-Solar metallicity of \feh=\Tmetal. 

\begin{figure}
    \centering
    \includegraphics[scale=0.5]{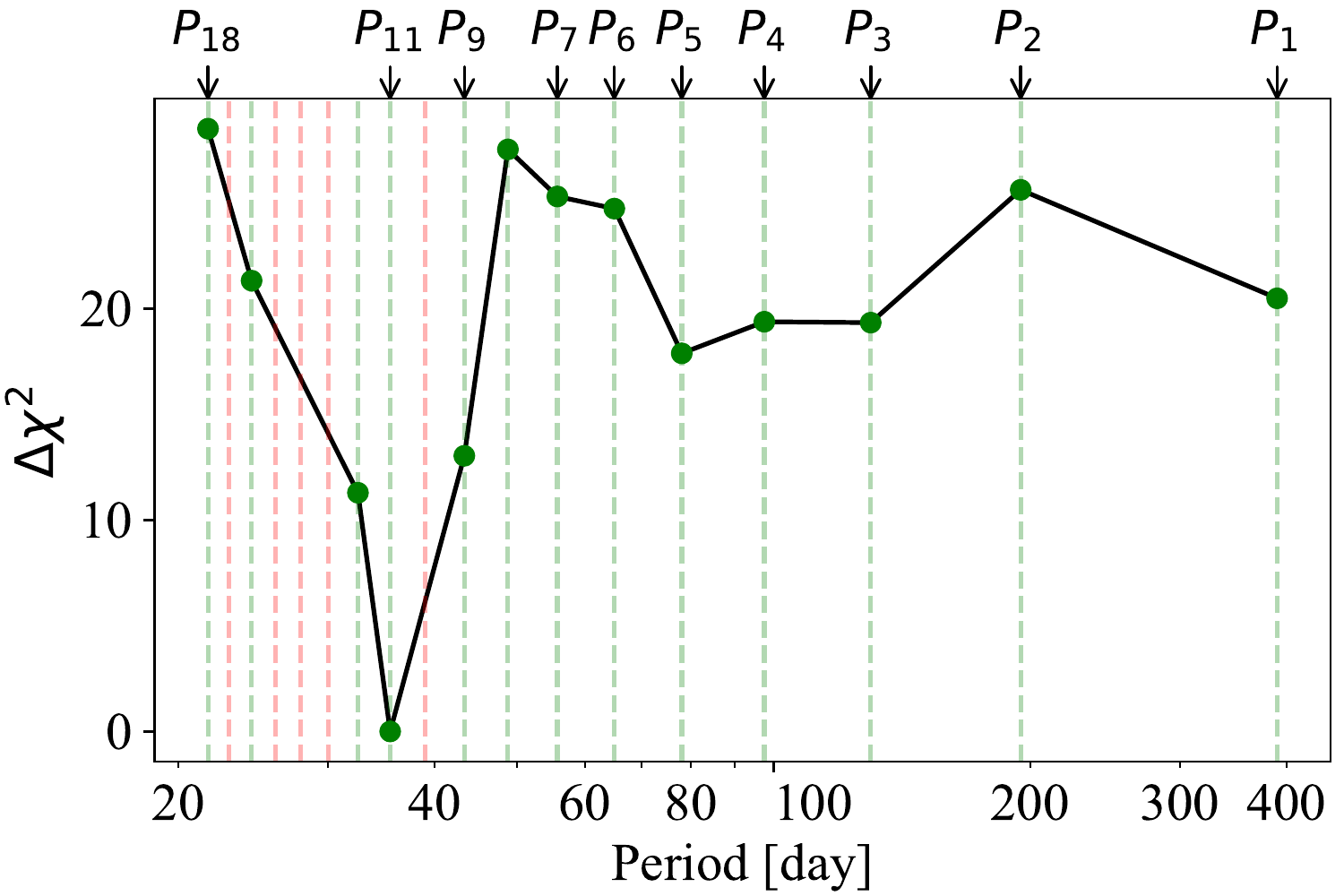}
    \caption{The relative goodness-of-fit for different candidate orbital periods from our joint fits to the \tess\ and \ngts\ transit detections and radial velocity measurements of \Tstar\ (described in Section~\ref{sec:combined}). Priors were used to constrain each fit to orbital periods around those allowed by the \tess\ and \ngts\ photometry (green dashed lines). Orbital periods excluded by \ngts\ observations are marked in red (see Section~\ref{sec:aliases}). Our best fit is to $P_{11}$ ($=\TdelTshort/11 = \Tperiodshort$\,days).}
    \label{fig:Figure_2}
\end{figure}

\subsection{Orbital period and planet properties}
\label{sec:combined}
In order to determine the true orbital period of \Tplanet, as well as its physical properties, we carried out joint fits of our transit light curves (\tess\ and \ngts) and radial velocity data (\harps\ and \feros) at each of the candidate orbital periods allowed by the photometry (see Section~\ref{sec:aliases} and Figure~\ref{fig:Figure_2}). 

A perfect minimisation algorithm would
%As Sam says, there is naturally a chi2 minimum at each of the orbital period aliases allowed by the photometry. For periods in between, the chi2 will be terrible because the transits will not fit. 
%If we had a perfect algorithm we could just allow it to 
explore the whole parameter space in a single fit, finding each of the thirteen orbital periods allowed by the photometry for itself. However, the $\chi^2$ minima are narrow and widely spaced, and this approach is not efficient in practice. Instead, we explored each minimum with a separate MCMC run, using a uniform prior that selected the period alias in question, but did not define the period itself. 
%and that won’t work in practice. So instead we found all of the allowed periods first (section 2.3) and then explored each of them with separate emcee runs - using a period prior that selected the alias but didn’t define the period tightly. 
Figure~\ref{fig:Figure_2} shows the best $\chi^2$ from each of these 13 runs, mapping the depth of each $\chi^2$ minimum.
%of these "individual fits”, but effectively they are a single fit where we have hoped emcee to explore each of the minima in turn. Note also that since the model and free parameters are identical for these different emcee runs, the delta-chi2 is actually identical to delta-BIC. 

Our model has 17 free parameters, with best-fitting values presented in Table~\ref{tab:Table1}, and the best-fitting model shown in Figure~\ref{fig:Figure_1}. In addition to the orbital period, $P$, these include $T_0$, the \tess\ transit epoch, $R_*/a$, $R_p/R_*$, and the transit impact parameter, $b$, where $R_*$ is the radius of the star, $R_p$ is the radius of the planet, and $a$ is the orbital separation. For the \tess\ and \ngts\ transit data we also fit for the out-of-transit magnitude offsets, as well as the decorrelated limb-darkening parameters $h_1$ and $h_2$ from \citet{2018A&A...616A..39M}. We employed Gaussian priors on $h_1$ and $h_2$ for each instrument interpolated from \citet{2018A&A...616A..39M} using stellar atmospheric parameters from Section~\ref{sec:specpars}. These priors dominate the posterior distributions for $h_1$ and $h_2$. In our initial fits we found that the \tess\ transit was slightly shallower than the \ngts\ transit, which we ascribe to residual blending with the neighbouring stars mentioned in Section~\ref{sec:tess}. We therefore introduced a parameter to allow for third light in the \tess\ light curve, $l_{3,\tess}$. Instead of fitting directly for eccentricity, $e$, and the argument of periastron, $\omega$, we use the decorrelated parameters $\sqrt{e}\sin \omega$ and $\sqrt{e}\cos \omega$, which perform better at low eccentricities \citep{2006ApJ...642..505F}. Our model also includes the radial velocity semi-amplitude of the host star, $K_*$, and the system velocity, $\gamma$. 
%To allow for absolute calibration uncertainties between \harps\ and \feros\ 

To account for different zero-points between \harps\ and \feros\ we also include a radial velocity offset for the \feros\ data with respect to \harps, $\Delta \gamma_{\rm FEROS}$.  To allow for astrophysical and/or instrumental noise not included in the formal radial velocity uncertainties we add a radial velocity jitter term in quadrature to the formal radial velocity uncertainties.  To obtain a spectroscopic reduced $\chi^2$ of unity, jitter terms of 5.5\,\ms\ for \harps\ and 19.0\,\ms\ for \feros\ are required. These jitter terms are in line with expectations for these instruments and a star of this brightness \citep[e.g.][]{Raynard18,2019AJ....157...55H,2020AJ....159..173H}. The bisector span shows no significant correlation with radial velocity.

We used \textsc{emcee} \citep{2013PASP..125..306F} to explore the parameter space around each candidate orbital period using the likelihood function $\mathcal{L}(\textbf{d}|\textbf{m}) = \exp (-\chi^2/2)$. We initiated 36 Markov chains and generated 100000 trial steps. We discarded the first 50000 as burn-in and visually confirmed the sampler had converged. Median parameter values for the best-fitting model at $P_{11} = \Tperiodshort$\,days are reported in Table~\ref{tab:Table1}, and the uncertainties are estimated using the 16$^{\rm th}$ and $84^{\rm th}$ percentiles of the cumulative posterior distributions.

Figure~\ref{fig:Figure_2} shows how the best-fitting $\chi^2$ depends on the orbital period alias. Our best fit is to $P_{11} = \Tperiodshort$\,days, and the next best fitting candidate period ($P_{12}$) has  $\Delta\chi^2=11.3$, which makes it more than 250 times less probable. 
It is important to understand that the fits to the different period aliases in Fig.\,\ref{fig:Figure_2} are effectively a single fit (with identical model and free parameters) exploring a complex $\chi^2$ space with multiple widely-spaced minima (at the periods allowed by the photometry in Sect.\,\ref{sec:aliases}). The $\Delta\chi^2$ plotted is identical to $\Delta\rm BIC$ (Bayesian information criterion).

\begin{figure}
    \centering
    \includegraphics[width = 0.46\textwidth]{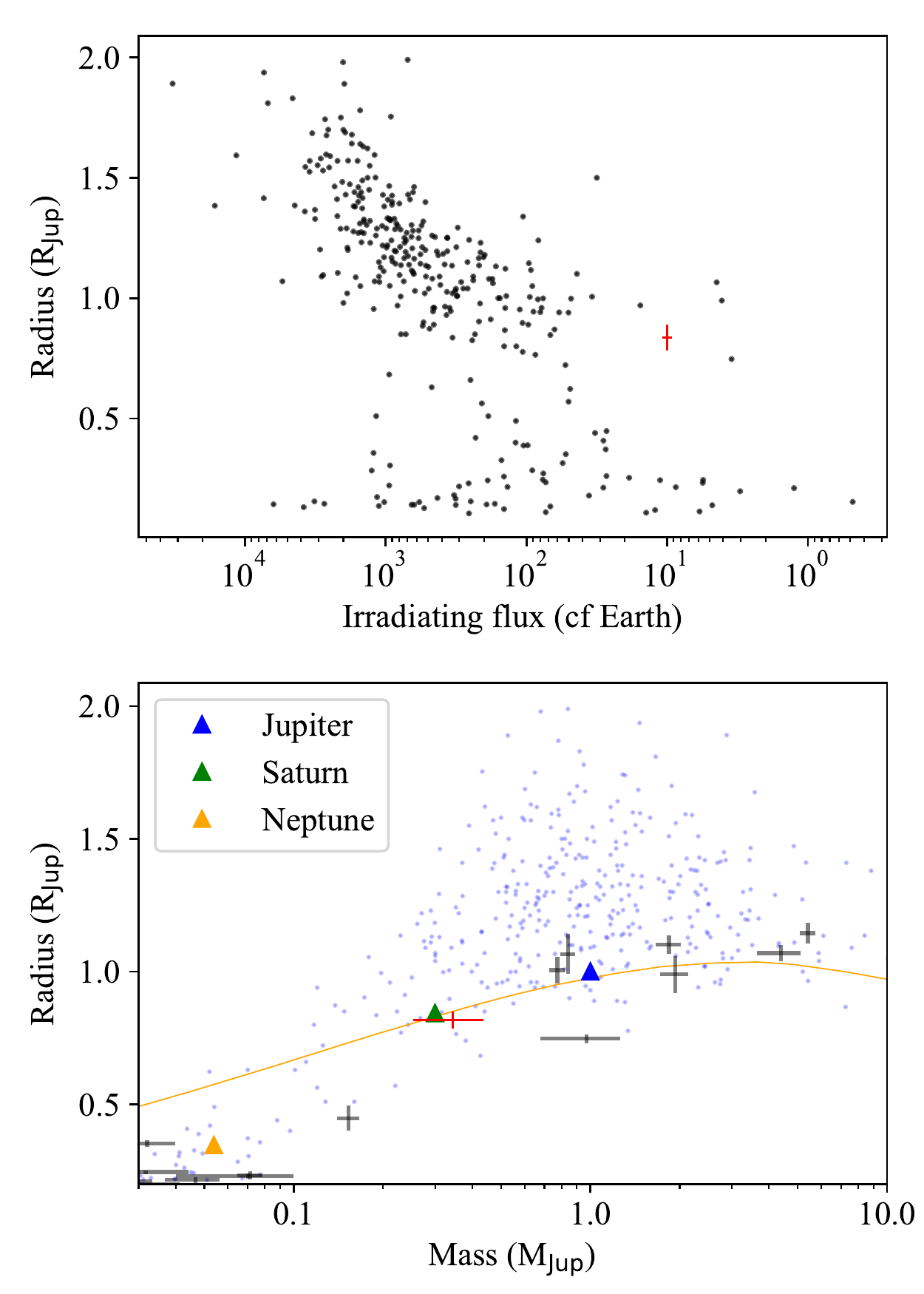}
    \caption{Upper panel: Radius-irradiation diagram of well-characterised transiting exoplanets (mass to better than 50\% precision and radius to better than 20\%; exoplanetarchive.ipac.caltech.edu, accessed 04/30/2020). \Tplanet\ is indicated in red. 
    Lower panel: Mass-radius diagram for giant exoplanets from the same sample. Planets with orbital periods longer than 30\, days are plotted in black, those with periods shorter than 30\,days in blue. \Tplanet\ is again represented with a red star and solar system giants are indicated with triangles. 
    The orange line shows a theoretical mass-radius relation for a cold hydrogen/helium exoplanet \citep{2007ApJ...669.1279S}.}
    \label{fig:Figure_3}
\end{figure}

Our best fitting model is shown in Figure~\ref{fig:Figure_1}, and the corresponding parameters are listed in Table~\ref{tab:Table1}. This shows \Tplanet\ to be an exoplanet with mass and radius very similar to Saturn (see Figure~\ref{fig:Figure_3}). 

\section{Discussion and conclusions}

Following-up a single-transit event identified in our \tess\ light curve of the K-dwarf \Tstar, we have detected a second transit event with \ngts. Combined with radial velocity follow-up with \coralie, \harps, and \feros, these transit detections show that the companion object is an exoplanet with mass and radius similar to Saturn. The planet has a \Tperiodshort\,day orbit that is marginally eccentric. \Tplanet\ is the first exoplanet to be confirmed after initially being identified as a \tess\ single-transit event. At our request, \Tplanet\ has also been designated \TTplanet.

Figure~\ref{fig:Figure_3} puts \Tplanet\ into context with the wider population of well-characterised exoplanets. The planet radius is not inflated compared to the solar system giants, as expected for a wide-separation planet \citep{Laughlin11,Thorngren18,Sestovic18}, and its composition is consistent with that of Saturn. \Tplanet\ has a surface gravity, $g_p$, of \Tpgrav\,\mss, calculated using the method of \citet{2007MNRAS.379L..11S}.  
Assuming an albedo similar to Saturn \citep[0.342; ][]{1983Icar...53..262H}, we calculate an equilibrium temperature of only \Teq\,K.  Since the host star is a K-dwarf, this equilibrium temperature is lower even than some longer-period planets with F- and G-type hosts.  In fact \Tplanet\ is one of just a handful of transiting gas giant planets with an equilibrium temperature $<500$\,K.  This makes \Tplanet\ an interesting prospect for transmission spectroscopy to study an atmosphere that is much cooler than the typically-studied hot Jupiters, which have equilibrium temperatures of 1000--2500\,K \citep[e.g.][]{Sing16}.  Assuming a Saturn-like atmospheric composition ($2.07\, \rm g \, \rm mole^{-1}$)\footnote{https://nssdc.gsfc.nasa.gov} we estimate an atmospheric scale height, $H_p$, of \TpH\,km, which may make it a viable target for transmission spectroscopy. The long baseline between the two transit detections already provides good precision for predicting future transits. For instance, the uncertainty in the timing of transit events in 2025 is only 28 minutes. 

It is important to note that we have been able to determine the mass and radius of this relatively long-period exoplanet with a very modest number of radial velocity measurements (nine with \harps\ and six with \feros). The detection of the second transit with NGTS was crucial in tightly constraining the possible orbital periods, and this serves to demonstrate the value of intense photometric monitoring in following-up single transit events. Without this second transit detection we would have required many more radial velocity measurements in order to confirm the planet, determine its orbital period, and measure its mass \citep[e.g. ][]{2020arXiv200310319D}. The strategy of large investments of photometric follow-up with instruments such as \ngts\ thereby allows efficient confirmation of single-transit events without adding to the considerable pressure on high-precision radial velocity instruments. 
This highlights the power of high-precision ground-based photometric facilities in  revealing longer period transiting exoplanets that \tess\ alone cannot discover.

\acknowledgments

The \ngts\ facility is operated by the consortium institutes with support from the UK Science and Technology Facilities Council (STFC) under projects ST/M001962/1 and ST/S002642/1. 

This paper includes data collected with the TESS mission, obtained from the MAST data archive at the Space Telescope Science Institute (STScI). Funding for the TESS mission is provided by the NASA Explorer Program. STScI is operated by the Association of Universities for Research in Astronomy, Inc., under NASA contract NAS 5–26555.

Based on observations made with ESO Telescopes at the La Silla Paranal Observatory under programme IDs $0104.C-0413$ (PI RB), $0104.C-0588$ (PI FB), Opticon:2019A/037 (PI DB), and CNTAC: $0104.A-9012$ (PI JIV).

The contributions at the University of Warwick by PJW, RGW, DLP, DJA, DRA, SG, and TL have been supported by STFC through consolidated grants ST/L000733/1 and ST/P000495/1. DJA acknowledges support from the STFC via an Ernest Rutherford Fellowship (ST/R00384X/1).
Contributions at the University of Geneva by FB, LN, ML, OT, and SU were carried out within the framework of the National Centre for Competence in Research ``PlanetS'' supported by the Swiss National Science Foundation (SNSF).
The contributions at the University of Leicester by MRG and MRB have been supported by STFC through consolidated grant ST/N000757/1.  SLC acknowledges support from the STFC via an Ernest Rutherford Fellowship (ST/R003726/1).
This project has received funding from the European Union's Horizon 2020 research and innovation programme under grant agreement No 730890.
AJ, RB and MH acknowledge support from project IC120009 ``Millennium Institute of Astrophysics (MAS)'' of the Millenium Science Initiative, Chilean Ministry of
Economy. AJ acknowledges additional support from FONDECYT project 1171208. RB acknowledges support from FONDECYT Post-doctoral Fellowship Project 3180246.
JSJ is supported by funding from Fondecyt through grant 1161218 and partial support from CATA-Basal (PB06, Conicyt).
MNG acknowledges support from the Juan Carlos Torres Fellowship.
ACh acknowledges the support of the DFG priority program SPP 1992 ``Exploring the Diversity of Extrasolar Planets'' (RA 714/13-1).
JIV acknowledges support of CONICYT-PFCHA/Doctorado Nacional-21191829.
EG gratefully acknowledges support from the David and Claudia Harding Foundation in the form of a Winton Exoplanet Fellowship.
TH acknowledges support from the European Research Council under the
Horizon 2020 Framework Program via the ERC Advanced Grant Origins 83 24 28.
This research has made use of NASA's Astrophysics Data System Bibliographic Services and the SIMBAD database, operated at CDS, Strasbourg, France. This research made use of Astropy,\footnote{http://www.astropy.org} a community-developed core Python package for Astronomy \citep{2018AJ....156..123T}.
\bibliography{references}{}
\bibliographystyle{aasjournal}

\end{document}